\documentclass[fleqn,usenatbib]{mnras}
\usepackage{newtxtext,newtxmath}
\usepackage[T1]{fontenc}
\usepackage{epsfig}
\usepackage[]{empheq}
\usepackage{natbib,ifthen}
\usepackage{wasysym}
\usepackage{mathrsfs}
\usepackage{color}
\usepackage{bbold}
\usepackage{mathrsfs}
\usepackage{csvsimple}
\usepackage[utf8]{inputenc} % allow utf-8 input
\usepackage{url}            % simple URL typesetting
\usepackage{booktabs}       % professional-quality tables
\usepackage{amsfonts}       % blackboard math symbols
\usepackage{nicefrac}       % compact symbols for 1/2, etc.
\usepackage{microtype}      % microtypography
\usepackage{amsmath}
\usepackage{graphicx}
\usepackage{wrapfig}

\usepackage{xcolor}
\usepackage{float}
\newcommand{\FT}{\mathcal{F}}

\title{Matched filtering with non-Gaussian noise for planet transit detections}

\author[J. Robnik and U. Seljak]{
Jakob Robnik,$^{1}$\thanks{E-mail: jakob\_robnik@berkeley.edu}
Uro\v{s} Seljak,$^{2, 3}$
\\
$^{1}$Department of Physics, ETH-H\"onggerberg, 8093 Z\"urich, Switzerland\\
$^{2}$Department of Physics,
University of California, Berkeley, CA 94720, USA\\
$^{3}$Lawrence Berkeley National Laboratory, 1 Cyclotron Road, Berkeley, CA
93720, USA
}

\pubyear{2020}

\begin{document}
%\label{firstpage}
\maketitle

\begin{abstract}
    We develop a method for planet detection in transit data, which is based on the Matched Filter technique, combined with the Gaussianization of the noise outliers. The method is based on Fourier transforms and is as fast as the existing methods for planet searches. The Gaussinized Matched Filter (GMF) method significantly outperforms the standard baseline methods in terms of the false positive rate, enabling planet detections at up to 30\% lower transit amplitudes. Moreover, the method extracts all the main planet transit parameters, amplitude, period, phase, and duration. By comparison to the state of the art Gaussian Process methods on both simulations and real data we show that all the transit parameters are determined with an optimal accuracy (no bias and minimum variance), meaning that the GMF method can be used both for the initial planet detection and the follow-up planet parameter analysis.  
\end{abstract}


\section{Introduction}
\label{se:intro}
Exoplanet detection using transits has 
become the leading method to detect new 
planets and determine their demographics.
Kepler Space Telescope \citep{KeplerSpaceTelescope} photometrically measured flux from about 200,000 stars, with over 4000 confirmed planets \citep{nasaExoplanetArchive}.
One of the primary goals of such studies 
is to determine the 
planet demographics, which is the planet occurrence 
rate as a function of its parameters, such as 
the radius of the planet and the distance from the star. This relies heavily on our ability to estimate how reliable these candidates are \citep{ImportanceForPlanetPopulation}. 
There are several origins of a false positive \citep{robovetter}: 
a planet can be mimicked by an eclipsing binary star, a single or multiple outlier noise event, a fluctuation of host star's brightness, a sudden instrumental drop, an event in an off-set star (a star in the same field of view
that has no physical association with a given star) \citep{false_positive}, etc. 
\par
For small planets far from the star distinguishing them from false signals is difficult, as such planets are close to or below the detection limit of the Kepler Space Telescope. One prominent such group are habitable zone Earth-like planets. Estimates for their occurrence rate varies wildly, with 95\% confidence 
interval covering more than one order of magnitude \citep{EarthLikePlanetsOccurence}. This has implications for the prospects of follow-up 
missions such as the proposed Luvoir \citep{luvoir}, where the 
predictions for the number of spectroscopically detectable habitable zone 
planets also varies by a similar amount. 

A traditional Kepler approach towards false positives is to perform a series of tests, each designed to target a specific group of false positives, and eliminate them if a candidate does not pass these individual tests. Those tests are, however, binary, meaning a candidate is 
either rejected or not, which 
is a rather crude approach when 
on the detection limit. A 
likelihood analysis or a Bayesian evidence analysis 
would be more informative for the subsequent hierarchical 
analysis. Also, some of the cuts are rather heuristic, such as checking for the proper shape of the transit by calculating a metric distance (LPP metric) from the known planet shapes \citep{robovetter}.
\par
The goal of this paper is to develop a new and 
independent planet detection pipeline, which 
is near optimal, fast, and provides sufficient 
statistical information for downstream tasks such as planet demographics. 
Our goal is to develop a rigorous analysis of the stellar variability and outlier false positives. 
{
Various Gaussian Process (GP) based methods have been developed that can be used to model stellar variability \citep{celerite,FourierGP} and determine how likely it is that a given candidate is caused by stellar variability.
}
These methods are however computationally expensive, so that they can only be used once a good candidate has been found and an initial estimate on its parameters is known. They must be combined with a simplified analysis of an actual planet search, where a simplified model for the stellar variability and planet transits is assumed. We will show that this is a crude assumption and results in higher significance of false positives, and therefore a loss of many real planet candidates, already at this initial stage of analysis. 
\par
We will present an alternative approach that is as fast as the simplified analysis currently used, but with the near optimal performance of the full Gaussian process analysis. It is therefore applicable to the complete Kepler data set with no need for the secondary GP analysis. The general idea behind our method is to use the Fourier based GP 
\citep{FourierGP}, which 
describes the stellar variability as a frequency dependent noise. This naturally connects to the matched 
filtering technique, which Fourier transforms the planet 
transit signal template(s) and performs signal to noise weighting in Fourier space first.  Afterwards one looks
for the highest peaks in its 
inverse Fourier transform. {Matched} filtering can be
shown to be optimal under the 
assumption of Gaussian noise, 
and is the 
method of choice in many statistical analyses, such 
as LIGO gravity wave signal detection \citep{LIGO}.
\par
Kepler data noise is not Gaussian, and noise outliers must be dealt with, otherwise they can lead to 
a significant increase in the false positive rate. 
Here we develop a Gaussianization transformation 
approach, which maps a signal with non-Gaussian power law tails in the Kepler data to a Gaussian. Specifically, we take advantage of the 
uncorrelated nature of noise to develop 
a method where this procedure does not change the 
planetary transit signal component. We will show that 
this approach eliminates the outlier false positives, and gives 
superior results to the alternatives such as robust statistics or outlier elimination \citep{outlierRejection}. 

\section{Noise Gaussianization transformation} \label{sec:gaussianization}

Here we first review the key results of \citet{FourierGP}. In general we write the 
data model
$d(t)$ as a sum of a transit signal $s(t)$,  
noise $n(t)$ and stellar variability $y(t)$. Here we first 
discuss the noise and the stellar variability, 
which can be viewed as a
correlated noise, so we  
assume
there is no transit signal. Stellar variability is correlated, and assumed to be Gaussian, which has been shown to be a good assumption \citep{FourierGP}. Noise is uncorrelated but non-Gaussian, distributed according to the noise probability distribution $p(n)$. In the absence of planet signal we assume to have a stationary, time ordered and equally spaced data $d_i = d(n \Delta t)$ for $n = 0, 1, 2, \dots, N-1$. 
\par
If we assume stationarity of the signal, the correlations depend only on the time difference between the points, and the GP kernel also depends only 
on their relative separation. Stellar variations could also be described with a
non-stationary form, but this would require a 
significant increase in the complexity of the 
kernel which we want to avoid. Later we will however 
describe the non-stationary generalization due 
to the gaps in the data. 
To describe 
a stationary kernel of a uniform time series the 
most general approach is to use the Fourier 
basis and describe the GP kernel using the 
power spectrum \citep{FourierGP}. 
A Fourier transform
\begin{align}
\tilde{y}_k= \FT\{y\}_k &= \frac{1}{\sqrt{N}} \sum_{n=1}^N y_n e^{i \omega_k n \Delta t} \\ \nonumber
y_n= \FT^{-1}\{\tilde{y}\}_n &= \frac{1}{\sqrt{N}} \sum_{k=1}^{N} \tilde{y}_k e^{- i \omega_n k \Delta t}.
\end{align}
introduces a new basis in which the covariance matrix is diagonal and can be 
described with the power spectrum, and the Fourier modes $\FT\{y\}_k$ are uncorrelated. We denoted $\omega_k = 2 \pi k / N$  .
\par
If on the other hand the data is uncorrelated, but 
non-Gaussian, then we can Gaussianize it as 
a simple 1-d point-wise non-linear transformation,
\begin{equation}
    \psi_i = \psi^{(1d)}(n_i).
\end{equation}
This transformation can be obtained by mapping the 
cumulative distributions of the data to a Gaussian \citep{FourierGP}.

{
Here we want to apply to the Kepler data, which we assume is composed of correlated and nearly Gaussian stellar variability, added to an uncorrelated noise, containing non-Gaussian outliers. Gaussianizing thus requires identifying the stellar component of the data, subtracting it from the data, Gaussianizing the remaining uncorrelated part, adding the stellar component back and transforming to the Fourier basis:
\begin{equation} \label{eq: fullGaussianization}
    {\mathcal G}(y) = \FT \{\Psi(d - y) + y \},
\end{equation}
In the absence of the correlated structures, such as planet transits, $\Psi = \psi^{(1d)}$. In general it is an invertible nonlinear scalar function which is local in the sense that it depends only on $d_i$ and possibly on its neighbours. We will use this generalization when dealing with the correlated structures in the data, such as planet transits.
The correlated Gaussian component $y(t)$ can be extracted from the data with the Fourier Gaussian process \citep{FourierGP}. Note that this step does not add much to the cost of our pipeline because it is done only once and because GP is never used for planet search and parameter inference, where it would have to be iterated with the optimization of the planet parameters.
}

We choose a function $\Psi(n)$ to Gaussianize the noise probability distribution $p(d)$, which then becomes 
\begin{equation}\label{nll1}
p(d) =\exp\left\{-\frac{1}{2}\sum_{k=1}^N  \bigg( \frac{|{\mathcal G}_{k}(d)|^2}{P_k} +\ln 2\pi +\ln P_k \bigg)
\right\},
\end{equation}
where $P_k$ is the $k^{\text{th}}$ component of the power spectrum of $y$.
Here $||^2$ is a product of a complex mode with its 
complex conjugate, which equals adding the squares of
its real and imaginary components. 

\subsection{Uncorrelated non-Gaussian noise}
\begin{figure*}
\includegraphics[scale = 0.45]{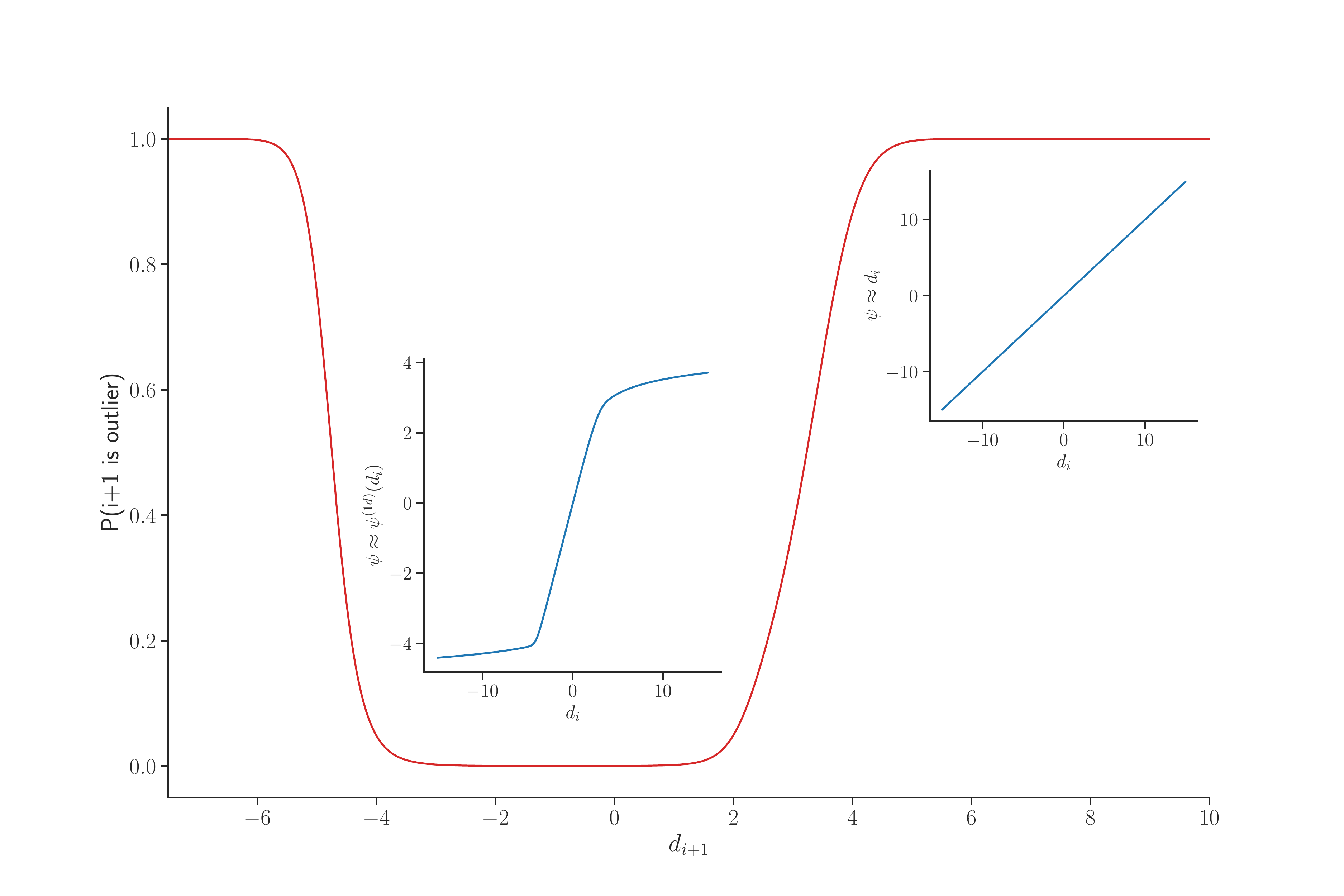} 
\caption{Gaussianization transformation at a point $i$ (blue line) depends on a flux $d_i$, as well as on it's neighbours $d_{i+1}$ and $d_{i-1}$. If negihbours are not outliers (central part of the red figure) a point $i$ is Gaussianized according to an 1-dimensional gaussianization, that is, outlier part of the distribution is mapped to the central Gaussian part. If however, neighbours are in the non-Gaussian part of the distribution (left and right parts of the red plot) then it is by far more likely that they are all a part of the correlated structure generated by a real transit and the Gaussianization acts as an identity.}\label{fig:gaussianization}
\end{figure*}

\begin{figure*}
\hspace*{-0.8cm}\includegraphics[scale = 0.45]{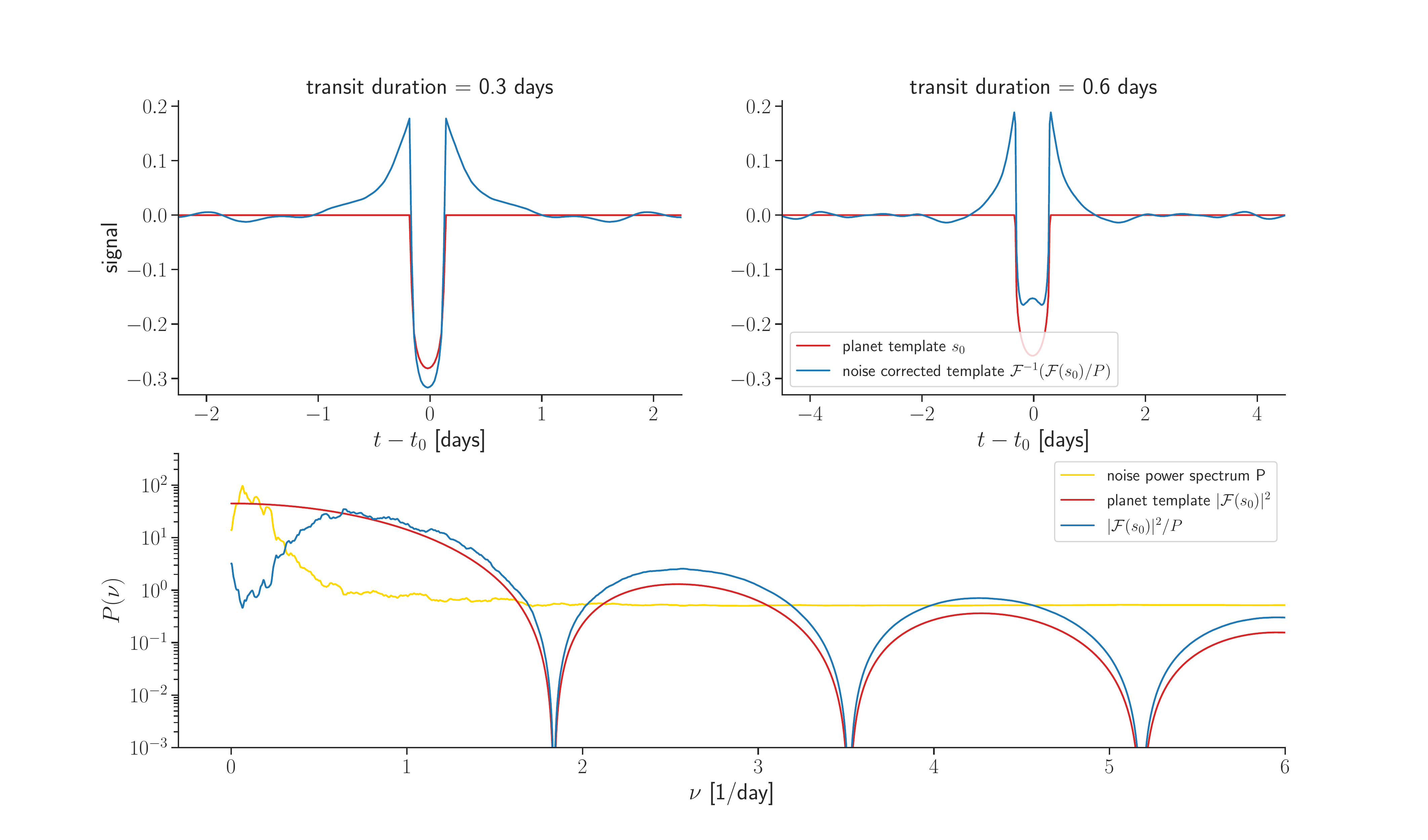} 
\caption{Top panels: a planet's transit signal $s_0$ (red) for the transit duration 0.3 days in the left panel and 0.6 days in the right panel. Match filtering with the frequency dependent noise as in Equation \ref{eq:snrdef} is equivalent to a convolution of a Gaussianized signal with the template, if we replace the original template $s_0(t)$ with the $\FT^{-1} \{ \FT \{s_0 \} / P_k\} (t)$ and normalize it. 
The matched filter template (blue) shows a partially compensated profile characteristic of a red noise power spectrum $P(\nu)$ with high power at low frequencies. The effect is more significant for longer transits, which can be more easily contaminated by the stellar variability. Bottom panel: noise power spectrum is taken from a realistic stellar variability analysis of star Kepler 90 (yellow), together with the 0.6 day planet transit template in the Fourier domain (red). Inverse noise weighting of the matched filter suppresses the low frequency components (blue) and leads to a compensated profile in time domain (top panels).}\label{fig: corrected template}
\end{figure*}

We need to determine the non-linear local $\psi(n)$ in equation \ref{eq: fullGaussianization}. Assuming uncorrelated noise, a given realization $n_i$ is distributed according to some probability density function $q$, independent of realizations at different times. Gaussianization ${\Psi}_i(n_i) = \psi^{(1d)}(n_i)$ will then also act point-wise and will satisfy 
\begin{equation}
P(\psi^{(1d)}(n_i)>X_N) = P(n_i > X_q),    
\end{equation}
where $X_N$ and $X_q$ are random variables distributed according to the normal and $q$ distributions. A unique function with the required property is
\begin{equation} \label{eq:1dGaussianization}
\psi^{(1d)} = CDF_N^{-1} \circ CDF_q   
\end{equation}
where $CDF_N^{-1}$ is an inverse of the cumulative density function of the normal distribution and $CDF_q$ is cumulative density function corresponding to $q$. 
\par
In our application noise is distributed normally except for the outliers. Outliers have been shown to be 
uncorrelated in Kepler data \citep{FourierGP}, so we can write for one data point $n_i$
\begin{equation} \label{eq:pdfNoise}
q(n_i) =  (1 - a) \, N(n_i) \, + \, a \; NCT(n_i),    
\end{equation}
where $N$ is a Gaussian distribution that is defined
by zero mean and variance that by Parseval's theorem is the sum over 
all power spectrum components, $NCT$ is a distribution modelling outliers and $a$ is a probability that a given realization is an outlier. 
%, with a 
%typical value for Kepler data of order $10^{-3}$. 
As shown in \citet{FourierGP}, the outlier probability density function can be modeled well with a non central t-distribution, and we determine its parameters and $a$ by a fit to the data 
PDF. For small amplitude $y$ the gaussian contribution dominates, as $a$ is typically a very small number of order $10^{-3}$. 
For large $y$ the $NCT$ contribution dominates as it decays only as a power law, in contrast to the Gaussian 
which decays very rapidly. %In practice if the star variability is slightly non-Gaussian it is also partially Gaussianized by this procedure. However, 
Correlated 
stellar variability never produces a large outlier signal, so it is not affected by the Gaussianization. 

\subsection{Filtering of correlated structures} \label{sec:filtercorr}
So far we assumed $d(t)$ to be a sum of noise and stellar variability, but in reality it also contains correlated structures such as planet transits, which may depart significantly from the average of the flux. We do not want this signal to be affected by the Gaussianization, i.e. we do not want the deep transit signatures to be mistaken for the noise outliers and have their depth reduced. 
Since outlier noise is uncorrelated while real transit is not 
we thus require that Gaussianization  $\psi^{1d}(d_i)$ to be identity if it acts on a correlated structure, and acts as a 1-d Gaussianization $\psi^{1d}(d_i)$ on an uncorrelated outlier. We write a full Gaussianization at a point $i$ as a mixture
\begin{equation} \label{eq: 3dGaussianization}
    \Psi_i = \mathscr{P} \, d_i + (1-\mathscr{P}) \, \psi^{1d} (d_i).
\end{equation}
$\mathscr{P}$ can be arbitrary if $\vert d_i \vert$ is small as $\psi^{1d} (d_i) = d_i$ then, and $\mathscr{P}$ cancels out. On the other hand if $\vert d_i \vert \gg 0$, $\mathscr{P}$ should be 1 if point $i$ is a part of the correlated structure and 0 if it is not. A simple but effective choice is
\begin{equation}
    \mathscr{P} (d_i| d_{i-1}, d_{i+1}) = P(\text{i+1 or i-1 is an outlier } \vert \,  d_{i-1}, \, d_{i+1} ),
\end{equation}
where $P$ is probability for an outlier that is calculated under the assumption that there are no correlated structures. We show that such $\mathscr{P}$ satisfies the required properties: 
\par
\begin{itemize}
    \item a priori probability of point $i$ being an outlier is $a$. If $i$ is not a part of a correlated structure then its neighbours are also outliers with probability $a$. Probabilities are independent, since we assume noise outliers are not correlated. A probability that $i$ and one of its neighbours are both outliers is then $2a^2$, which can be well approximated to be zero for a typical value of $a$ of $10^{-3}$: we assume the probability of two neighbours both being an outlier is zero. Therefore $\vert d_i \vert \gg 0$ $\implies$ $i\pm1$ are not outliers $\implies$ $\mathscr{P} = 0$.
    \item on the other hand if $\vert d_i \vert \gg 0$ and point $i$ is a part of the correlated structure then also $\vert d_{i+1} \vert \gg 0$ or $\vert y_{i-1} \vert \gg 0$ and $\mathscr{P}$ is close to 1, as required. 
\end{itemize}
$\mathscr{P}$ is straightforward to compute:
\begin{align}
    \mathscr{P} &= 1- P(\text{i+1 and i-1 not outliers } \vert \,  d_{i-1}, \, d_{i+1} ) \\ \nonumber
    &= 1- P(\text{i+1 not outlier} \vert \,  d_{i+1} ) P(\text{i-1 not outlier } \vert \,  d_{i-1} )
\end{align}
where $P(\text{not outlier } \vert \, d)$ is computed by a simple application of the Bayes theorem:
\begin{equation}
    P(\text{not outlier } \vert \, d) = \frac{(1-a) N(d)}{(1-a) N(d) + a NCT(d)}.
\end{equation}
The Gaussianization of Equation \ref{eq: 3dGaussianization} preserves the correlated structures, such as planet transits or binary star transits, and reduces the noise outliers by mapping them to the Gaussian distribution, thus reducing their impact on the outlier false positives in the search for the planets. We will show this procedure is more effective than outlier 
rejection, and is easy to implement. 

\section{Matched filter}

Next we look for a signal $s$ in the data $d$ in the presence of the correlated non-Gaussian noise $y + n$, such as considered in the section \ref{sec:gaussianization}. The result will be a frequency dependent filter, which is a 
generalization of the Gaussian matched filter because of the Gaussianization we perform on the 
noise. In section \ref{sec:periodicTemplate} we will specialize to the localized periodic templates that can be used for the planet search, which will be further addressed in the next section.
\par
The signal has a template form of an event with a time profile
\begin{equation} \label{eq:template}
    s(t) = A \, s_0(t),
\end{equation}
where $A$ is an amplitude of the signal and $s_0$ is a template. Template depends on additional parameters such as the time of transit $t_0$ or transit duration. Shape of the template is for now assumed to be known, we will determine it's parameters by a template bank search method. An example of a signal is shown in the Figure \ref{fig: corrected template}.
It shows the effect a typical red power spectrum 
taken from a realistic stellar variability analysis of star Kepler 90 on the matched filter. If the power 
spectrum were white there would have been no effect, but 
for the red power spectrum the effect shows up as 
a partially compensated profile, which suppresses the low 
frequencies that are contaminated by the correlated noise and hence need to be filtered out. 
The effect is more significant for longer transits,
because the stellar variability has more power on longer 
timescales.

\par
We can write the data $d(t_i)$ as 
\begin{equation}
    d(t_i) = y(t_i) + n(t_i) + A s_0(t_i).
\end{equation}
By Gaussianizing the residuals under the 
assumptions of section \ref{sec:filtercorr} we obtain
\begin{align}
    \Psi_i(d-s) &= \Psi_i(d) - s_i, \\
    \mathcal{G}_k(d - s) &= \mathcal{G}_k(d) - \FT\{s\}_k
\end{align}
Equation \ref{nll1} becomes
\begin{equation}\label{eq:logp}
-2\ln p (y) = \sum_{k=1}^N  \frac{|{\mathcal G}_{k}(d)- A \FT\{s_0\}_k|^2}{P_k} + c
\end{equation}

where $c =\ln 2\pi +\ln P_k-2\ln J(d-s)$. The main advantage of the matched filter technique is that it can analyze the data for every possible value of $t_0$, by performing FFT based convolution of the data with the signal. This is complicated in the nonlinear case by the presence of the Jacobian term in $c$. In a typical Gaussianization application the Jacobian term is negligible, so we will drop this term.

At a given template there is a unique extremal value of the amplitude $\widehat{A}$, which can be obtained 
from the maximum likelihood, for which the derivative
\begin{equation}\label{zeroderivative}
    \frac{\partial \ln p}{\partial A} \bigg\rvert_{\widehat{A}} = 0.
\end{equation}
Taylor expanding the log-likelihood function around the optimal amplitude we obtain  
the variance of $A$
%\begin{equation}
%    -2 \log \frac{p (A)}{p(\widehat{A})} =  \frac{1}{2}\frac{\partial^ 2 (-2 \log p)}{\partial A ^2}\bigg\rvert_{\widehat{A}}  (A-\widehat{A})^2 \equiv  \frac{(A - \widehat{A})^2}{\sigma_A^2},
%\end{equation}
%gives a standard deviation of A:
\begin{equation}
    \sigma_A^{-2} (t_0) = -\frac{\partial ^2 \ln p}{\partial A ^2}\bigg\rvert_{\widehat{A}} .
    \label{var}
\end{equation}
Equating first derivative to zero gives
\begin{equation} \label{eq:amplitude}
\widehat{A}(t_0, s_0) = \frac{\sum_{k=1}^N {\mathcal G}_k^* \FT\{s_0\}_k /P_k}{\sum_{k=1}^N |\FT\{s_0\}_k|^2 / P_k} .
\end{equation}
Equation \ref{var} gives
\begin{equation} \label{eq:amplitudeSTD}
    \sigma_A^{-2} (t_0, s_0) = \sum_{k=1}^N |\FT\{s_0\}_k|^2 / P_k .
\end{equation}
A signal to noise for the event $s_0$ happening at time $t_0$ is then defined as the ratio of the signal to 
variance 
\begin{align}
    SNR(t_0) &= \frac{\widehat{A}(t_0)}{\sigma_A (t_0)} = \\
    &= \frac{\sum_{k=1}^N {\mathcal G}_k^* \FT\{s_0\}_k /P_k}{\bigg( \sum_{k=1}^N |\FT\{s_0\}_k|^2 / P_k \bigg) ^{1/2}}  \label{eq:snrdefGaps}\\
    &= \frac{\FT^{-1} \bigg{\{}{\mathcal G}_k^* \FT\{s_0\}_k /P_k \bigg{\}}}{\bigg( \sum_{k=1}^N |\FT\{s_0\}_k|^2 / P_k \bigg) ^{1/2}}
     \label{eq:snrdef},
\end{align}
{
where the last step is valid under the assumption that the template is stationary .
%, that is, if $s_0(t, t_0) \neq s_0(t - t_0)$. 
}
As expected from a matched filter, the $SNR$ is proportional to a convolution of the Gaussianized signal $\Psi(d)$ with the filter profile $s_0$, which is a multiplication of their corresponding Fourier transforms, followed by an inverse Fourier transform. Noise power modulates this convolution with the inverse noise weighting: the larger the noise power $P_k$, the less weight a given Fourier component contributes. The denominator term properly normalizes the $SNR$ by the expected signal of the matched filter profile, again inverse weighted by the noise power. 
\par
The computational complexity of evaluating the Equation \ref{eq:snrdef} for some value of $t_0$ is $O(N \log N)$. More importantly, evaluating it for all $t_0$ on a time lattice with a lattice spacing $\Delta t$ is still $O(N \log N)$ thanks to the Fast Fourier Transform. This is very useful for a search over the whole parameter space, which is required in the initial planet search where we 
do not know planet period, phase, or transit 
amplitude. 

{

Note however, that the template may not be stationary. For example, if there are gaps in the data, the template $s_0(t, t_0)$ has zeros where the data is missing and those zeros cannot be shifted. In this case the matched filter cannot be calculated by an inverse Fourier transform and the cost of evaluating Equation \ref{eq:snrdef} for all $t_0$ on a time lattice is $O(N^2 \log N)$. We will show in the next subsection how to simplify and avoid incurring this cost in the planet search. 
}

\subsection{Periodic template} \label{sec:periodicTemplate}
A special case of interest is a template containing multiple events which repeat with a period $P$ (not to be confused with the power spectrum $P_k$). The template has the form
\begin{equation} \label{eq:periodicTemplatedef}
 S_0(t, P, \phi) = \sum_{m \in I} s_0(t - m P - \phi)
\end{equation}
where $\phi$ is a phase, $I$ is a set of all integers $m$ for which the data at $m P + \phi$ is available and $s_0$ is a template of each individual event. An example of $S$ would be multiple transits of the planet. 
\par
In principle one would have to add $P$ as a parameter of the template and find it using a template bank approach. 
%Even worse, Equation \ref{eq:snrdefGaps} should be used due to the gaps in the data. 
However for the purpose of finding good planet candidates in the Kepler data a few simplifications can be made to make this faster.
\par
We will assume that:
\begin{itemize}
    \item Events $s_0$ are localized: overlap sums containing different events $\FT\{s_0(t)\}_k \FT\{s_0(t - P)\}_k^*$ are zero. This can be used to write the $SNR(P, \phi, S_0)$ in terms of the $SNR(t_0, s_0)$. This assumption is in principle problematic for the planets with short periods, but we verified that even for planets with a 3 day period this would result in only a 3 \% bias of the $SNR$.
    \item For each $m$ we assume that for all $t$ close to $\phi + m P$ either almost all data is available or almost all data is missing. By close we mean those $t$ that contribute most to the $SNR$ of the event. Then the template is approximately stationary and the Equation \ref{eq:snrdef} can be solved by the inverse FFT. This assumption fails only for the transits with partially missing data. We justify this assumption by noting that 
    most of the missing data is collected in the time gaps which are long compared to the transit time. 
\end{itemize} 
 Inserting Equation \ref{eq:periodicTemplatedef} into Equation \ref{eq:snrdef}, using linearity of the Fourier transform, and the above stated assumptions, we derive
\begin{equation} \label{eq:periodFolding}
 SNR(P, \phi, S_0) = \frac{1}{\sqrt{\vert I \vert}} \sum_m SNR(m P + \phi, s_0).
\end{equation}
where $SNR(P, \phi, S_0)$ is a joint signal to noise ratio of all periodic events with the period $P$ and phase $\phi$ found with the template $S_0(t, P, \phi)$. $SNR(t_0, s_0)$ is a $SNR$ of one individual event centered at $t_0$, found with the template $s_0(t, t_0)$.
Thus one can look for periodic events by first applying convolution \ref{eq:snrdef} to get $SNR(t_0, s_0)$, and then fold it at different periods using Equation \ref{eq:periodFolding}.

\begin{figure*}
    \centering
    \hspace*{-1.7cm}\includegraphics[scale = 0.4]{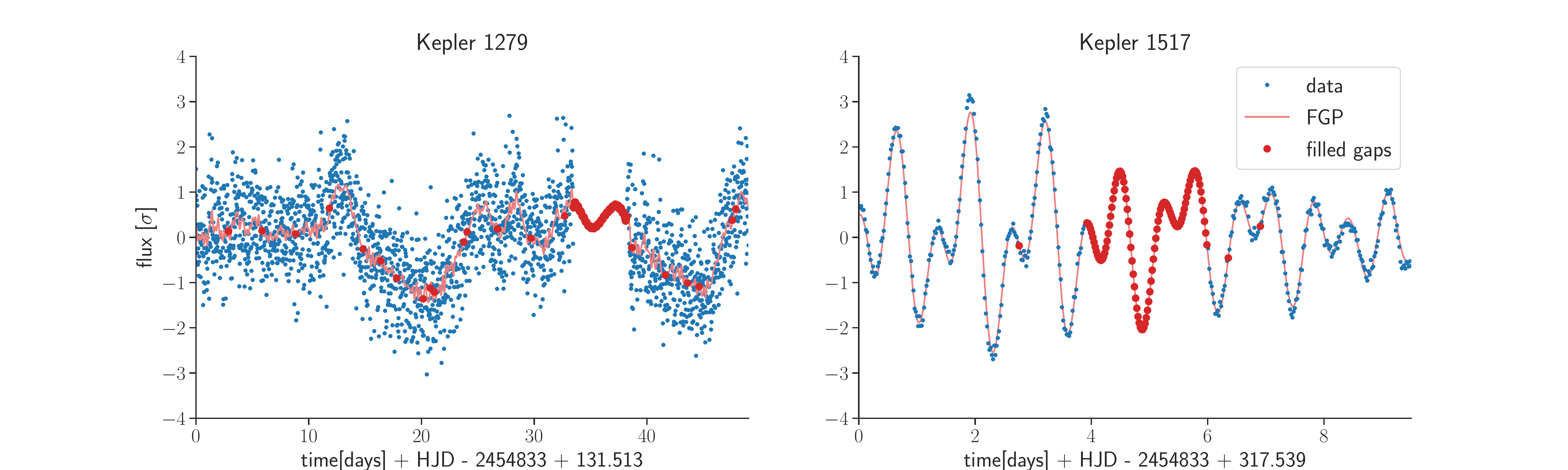}
    \caption{
    {
    Kepler data has gaps where no data is available. Data is missing in the form of either isolated data points or as larger gaps than can be as long as 3 months. We fill the gaps by fitting the Fourier Gaussian process to the data. We show a segment of the data for two stars, Kepler 1279 and Kepler 1517. Red dots are added to the data to fill the gaps.
    }
    }
    \label{fig: gap filling}
\end{figure*}

\section{Planet search} \label{sec: planet search}

We now apply the Gaussianization and matched filter (GMF) formalism to an example planet search in the Kepler data. We use the Kepler data processed through the Pre-search data conditioning module \citep{kepler_process}, which eliminates systematic instrumental errors. Specifically, 
we use PDCSAP flux, where long term trends have been eliminated. We normalize flux in different quarters as described in \citet{FourierGP}, to get an evenly spaced time series (except for gaps in the data), with unit variance of the Gaussian part of the distribution and zero average. We  Gaussianize the data with Equation \ref{eq: 3dGaussianization} to obtain a normally distributed, correlated noise, that may contain
correlated structures, such as planet transits.
\par
{ 
Kepler time streams have gaps where the data is not available. Filling this points with zeros is not advisable if the stellar flux is highly correlated on short time-scales (e.g. Kepler 1517 in Figure \ref{fig: gap filling}), as this is introducing a jump in the data which may trigger a false positive planet event. Instead, we use the Fourier GP \citep{FourierGP} to determine the correlated stellar component of the data and insert it in places where there are gaps. An example of gap filling is shown in Figure \ref{fig: gap filling}.
}
\par
We will first discuss the template form of the signal from the Equation \ref{eq:template}. It is modeled by the dimensionless template form $U$ and the transit duration $\tau$ as described in \citet{FourierGP, transitImplementation}:
\begin{equation}
    s_0(t-t_0) = U((t-t_0)/\tau, u_1, u_2).
\end{equation}
Limb darkening parameters $u_1$ and $u_2$ are a property of the star, their impact will be discussed in the Section \ref{sec:LimbDarkening}. $U$ depends weakly on the radius of the planet, which can be accounted for in an iterative manner, but we verified that its impact on the $SNR$ is less then one in a thousand, so we will ignore it.  In the planet search we will first adopt a template bank approach and search over the entire parameter space to find the best planet candidates, and then optimize with respect to the planet's parameters once we are close to the peak.
\par
For a matched filter analysis we first need the noise  power spectrum $P_k$, whose determination will be described in the Section \ref{sec:Pkestimation}.
Next we look for the planets which are significant enough to be detectable without the period folding (Section \ref{sec:largePlanets}). In this approach we do not lose information on the potential {transit timing deviations (TTV) and transit duration deviations (TDV)}, which can be used for planetary dynamics studies, for example planet-planet or planet-moon gravitational interactions. Also, if TTVs are large, a folded analysis would leave residuals, which we want to avoid. In the last stage (Section \ref{sec:smallPlanets}) we look for small planets, where period is added as a parameter, and a folded analysis is required to reach a sufficient signal to noise.
\begin{figure*}
\hspace*{-1.4cm}\includegraphics[scale = 0.4]{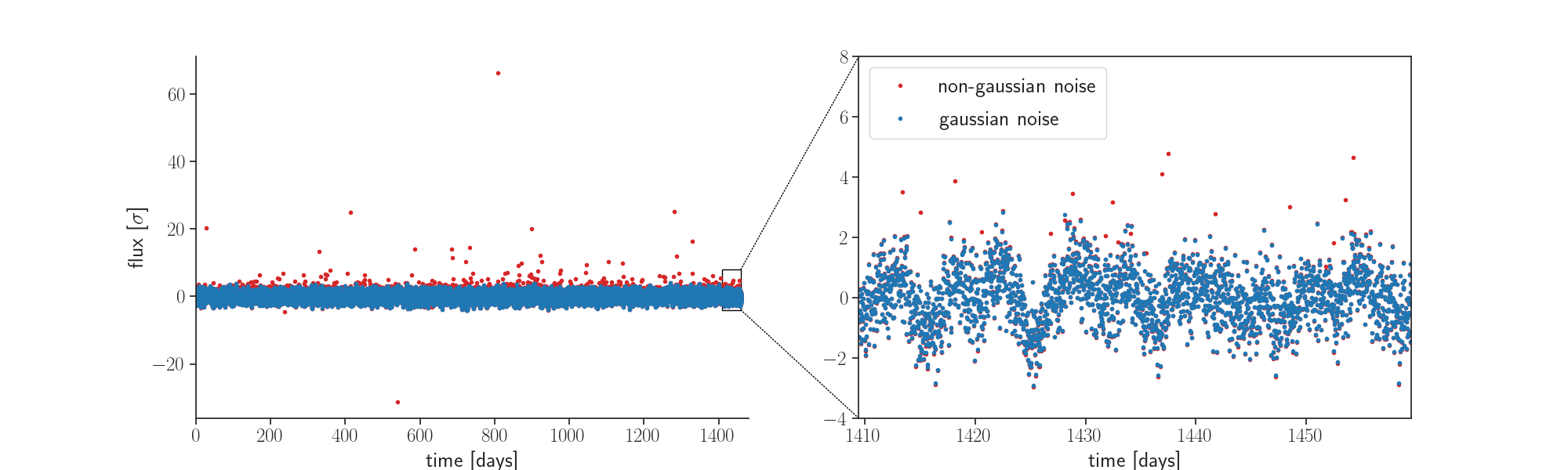}
\caption{ { Simulation of stellar background time series (blue) and added non-Gaussian outliers (red). Noise is correlated with a red power spectrum. In this example the power spectrum is extracted from the Kepler 90 data (shown in Figure \ref{fig: corrected template}). Right panel is zoomed part of the left panel, showing the correlations. This one realization of the stellar background time series that will be used to test the false positive rate in Figure \ref{fig:pvalues}}} \label{fig: example realization}
\end{figure*}
\subsection{Limb darkening parameters} \label{sec:LimbDarkening}
Stellar flux density over the stellar disk is modeled by a quadratic law in cosine of the angle spanned by the observer, center of the star and the point on the surface of the star, as proposed in \citet{simple_limb_darkening}. Coefficients of the polynomial $u_1$ and $u_2$ can only take values in the triangular region of the $\mathbb{R}^2$ set by the requirement that physical profiles must have positive flux density and that flux density is decreasing as the angle is increasing. In \citet{q1q2reparametrization} they give a convenient reparametrization $(u_1, \,u_2) \mapsto (q_1, \,q_2)$, such that physical region corresponds to $(q_1, \,q_2) \in [0, 1]^2$. Reparametrization also decreases correlations between the limb darkening coefficients so we use these coordinates.
\par
Limb darkening parameters can be calculated from the properties of the star: effective surface temperature, surface gravitational acceleration and metalicity \citep{u1u2fromstar}. As a result of the experimental error in the stellar parameters and systematic errors in the predictions, limb darkening coefficients are still only constrained to a small subspace of the $[0, 1]^2$ plane. Limb darkening parameters do not have a strong impact on the $SNR$ of the planet candidate as compared to period $P$, phase $\phi$ and transit duration $\tau$: theoretical predictions from the stellar properties can be used as an initial guess to find all of the planet candidates. Joint $SNR$ of all planet candidates can then be maximized with respect to the limb darkening parameters if needed.

\subsection{Noise power spectrum} \label{sec:Pkestimation}
Noise is composed of a white detector noise and a correlated component due to the stellar variations. Power spectrum of the stellar variations approaches zero at high frequencies, making the white noise component dominant. We assume the true noise power spectrum is a smooth function of frequency. After Fourier transforming the data and multiplying with the complex conjugate we perform a band power averaging to reduce the variance of individual bandpowers. At high frequencies we use wider bands, as power spectrum is roughly constant, and variations are mostly the standard fluctuations of a
Gaussian process. 
\par
Presence of the data gaps requires the iterative estimation of the power spectrum \citep{FourierGP} where in the $n$ -th step the just estimated power spectrum $P^{(n)}$ is used to simulate a flux series with gaps whose power spectrum $P_{sim}^{(n)}$is used to correct the estimate of the power spectrum in the next step $P^{(n+1)} = P + P^{(n)} - P_{sim}^{(n)}$, where $P$ is the estimate of the power spectrum obtained from the data. In practice a few steps are sufficient for the convergence.
\par
If a large planet signal is present in the data it affects the power spectrum at high frequencies due to its U-shape. Thus we first find the large planets using a first rough estimation of the power spectrum, eliminate these
large planets from the flux, and recalculate the 
power spectrum. We only do this once, as we find that an additional iteration on this process is not necessary.

\subsection{Search for the large planets} \label{sec:largePlanets}
By large planets we mean those planets whose individual events are significant on their own. First, using Equation \ref{eq:snrdef} we convolve the data with the template forms of different transit duration. We maximize over the transit duration first and then find peaks over the time of the transit. For the found candidates we then optimize the exact $SNR$ from Equation \ref{eq:snrdefGaps} which properly accounts for the gaps. These candidates are not forced to have equal transit 
time, and a subsequent analysis can be used to 
determine their TTVs. 

\subsection{Search for the small planets} \label{sec:smallPlanets}
Finding small planets is more challenging, specially those on the edge of detectability. It requires a search over their period, phase and transit duration. First, using Equation \ref{eq:snrdef} we convolve the data with template forms of different transit duration. Then using Equation \ref{eq:periodFolding} we search over the planet's parameter space. If the planet's orbit was circular and perfectly aligned with the line of sight the transit duration would be completely fixed by the period: $\tau_K = q P^{1/3}$, by the Kepler's third law. The proportionality constant is given by the stellar radius $R_*$ and the stellar mass $M_*$: $q = R_*(4 / \pi G M_*)^{1/3}$. Inclined and elliptical orbits allow for a different transit duration, but it is sufficient to first fix the transit duration to the $\tau_K (P)$, maximizing over $\phi$, and then only consider different $\tau$ for the highest SNR candidates. This yields a top candidate at a given period, after neglecting all the other candidates at the same period, knowing that there cannot be two planets at exactly the same period. Peaks in the thus found $SNR(P) = max_{\phi, \tau} SNR(P, \phi, \tau)$ correspond to the planet candidates. This 
procedure assures no real planets have been 
eliminated, but some of the candidates may be false positives. We eliminate next those
candidates that are caused by some confirmed planet with higher $SNR$, e.g. it's higher harmonics. Such false positives appear because it would be unnecessarily expensive to terminate the search every time a new candidate is found, since it would 
require to eliminate it from the data and start over, which can be expensive especially for the systems with many planets. Instead we find all the candidates first, and then sweep over the candidates, starting at the highest $SNR$, removing them from the data one by one, and identifying candidates lower on the list that still have $SNR$ above the threshold. This procedure 
ensures that the found planet candidates are  independent and not higher harmonics of higher $SNR$
candidates.

\section{Testing GMF on simulations and real data}
We now compare our GMF method with baseline methods used in the literature. We test the ability to extract the correct amplitude of the injected planet, provide the  smallest errors and accurately estimate it, and to produce as few false positives as possible. We show that GMF significantly outperforms the 
baselines in terms of the false positives, and at the same time is as accurate in reproducing the $SNR$ as the most sophisticated Gaussian process methods (which cannot even be used as a planet search engines due to their excessive computational cost).
\subsection{False positives test}

\begin{figure*}
\hspace*{-1.2cm}\includegraphics[scale = 0.35]{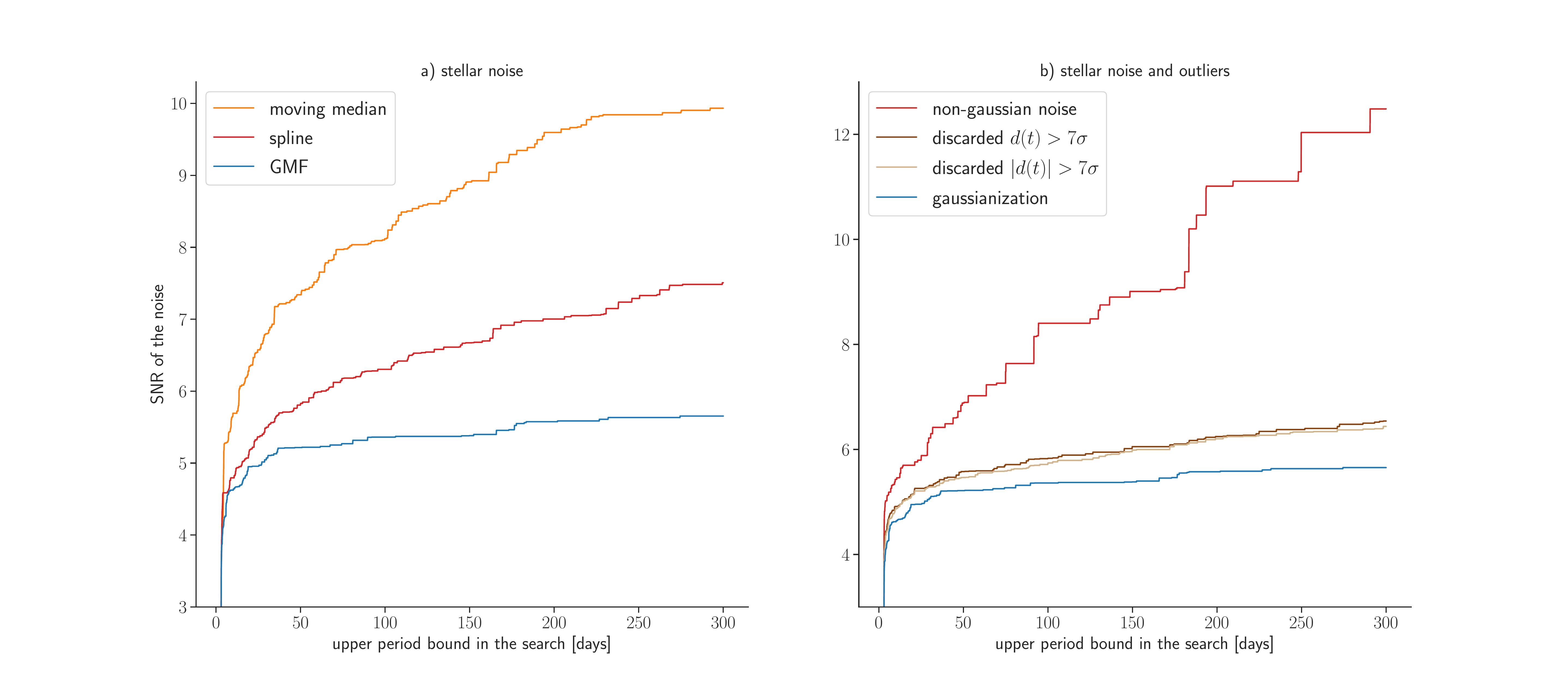} 
\caption{ Expected maximal $SNR$ event in the background only simulation as a function of the maximal period $T$, scanning over a period from 3 days to $T$. Maximum $SNR$ event in the search at each upper limit $T$ is computed for 30 noise realizations. We show the median over the realizations (50\% have higher false positive $SNR$). Left panel a): here the background is a Gaussian, correlated noise with a power spectrum taken from the Kepler 90 (example of a realization of such noise is shown in blue in Figure \ref{fig: example realization}). We show different methods to the planet search: moving median (orange) eliminates star flux by a moving median, spline (red) does this with a spline interpolation. They both eliminate the estimated stellar flux and proceed by using a matched filter with a flat noise power spectrum. GMF (blue) proposed in this paper does not eliminate the stellar flux but treats it as a correlated component of the noise. GMF has significantly lower median $SNR$ of false events. Right panel b): non-Gaussian outliers are added to the time series (example of a realization is shown in red in Figure \ref{fig: example realization}). We show the case where outliers are not accounted for (red), a cutoff of positive outliers (dark brown), cutoff of all outliers (light brown) and the Gaussianization (blue) In all cases we use our frequency dependent match filter which performed best in the case 
a). Gaussianization has the lowest $SNR$ of the false positive events, so it has the lowest false positive rate at a given $SNR$.} \label{fig:pvalues}
\end{figure*}

As our initial test we simulate a background-only time series. Stellar background is a Gaussian correlated noise, with the power spectrum resembling the power spectrum of Kepler 90 \citep{FourierGP}. A given realization is a random {sample from a Gaussian distribution}. Later we also add the noise outliers, such that each point is an outlier with some small probability $a$, independently of the other points, and is thus drawn from the non-Gaussian distribution (Section \ref{sec:gaussianization}). A realization of both processes is shown in Figure \ref{fig: example realization}.
\par
We search for the planet shaped transits in the background over periods in range between $T_{\rm min}=3$ days and $T_{\rm max}=300$ days, over all phases and fix a transit duration to the Kepler value $\tau_K$. We determine the maximal $SNR$ as a function of period $T$ cumulative from $T_{\rm min}$ to $T$.
We report the median over different realizations, so that 50\% of realizations have higher $SNR$. In Figure \ref{fig:pvalues} we show the median of the maximal SNR as a function of the period $T$, so this is the
highest $SNR$ between 3 days and $T$ range over which the search is performed.
\par
First we show the different methods of the planet search using the stellar background only. One common method is to approximate the stellar variability by spline fitting with spacing of nodes on a time scale which is larger than the transit duration \citep{spline}, and then subtract it from the time series. A matched filter with a flat noise power spectrum is then performed to find the planet candidates. Another approximation for the stellar variability is to take a moving median across bins which are longer than a typical transit duration \citep{foremanplanetsearch}. Wavelet based methods for the planet search \citep{planetSearchWavelets} similarly assume a separation of the time scale between the planet duration and the star variability. { In Figure \ref{fig:pvalues} we show that our method has significantly lower false positive rate than the spline fitting and the moving median.}
\par
Next we add outliers to the background. Here we use our best performed matched filter method to 
address the impact of noise outliers only. 
A standard practice \citep{kepler_process} is to introduce a cutoff on the outliers such that all points with a flux deviation from the mean larger than a cutoff are discarded from the series, here 
chosen to be 7 sigma. This is problematic for the negative outliers because  the excluded points may be from a planet signal.
In contrast, Gaussianization fully preserves the planet signatures. But even in the absence of this problem, such as in the Figure \ref{fig:pvalues}, cutoff method is not as good as the Gaussianization, because it must be made in the region where the Gaussian distribution is negligible, whereas Gaussianization also operates in the region below 7 sigma, where both the outlier distribution and the Gaussian distribution are important. Gaussianization 
procedure is optimal noise outlier suppression 
method by construction, and this is reflected in the 
reduced false positive rate.
\par
We note that positive outliers can produce false positive events in the presence of the stellar variation because the effective template is positive in some regions, see Figure \ref{fig: corrected template}. Positive outliers in the Kepler data are more prominent than the negative outliers which makes this effect comparable to the false positives from the negative outliers. This is not an artefact of match filtering with the stellar variability, and standard Kepler pipeline also encounters this problem \citep{kepler_process}. 

\begin{figure*}
\vspace*{-0.2cm}\hspace*{-1.2cm}\includegraphics[scale = 0.38]{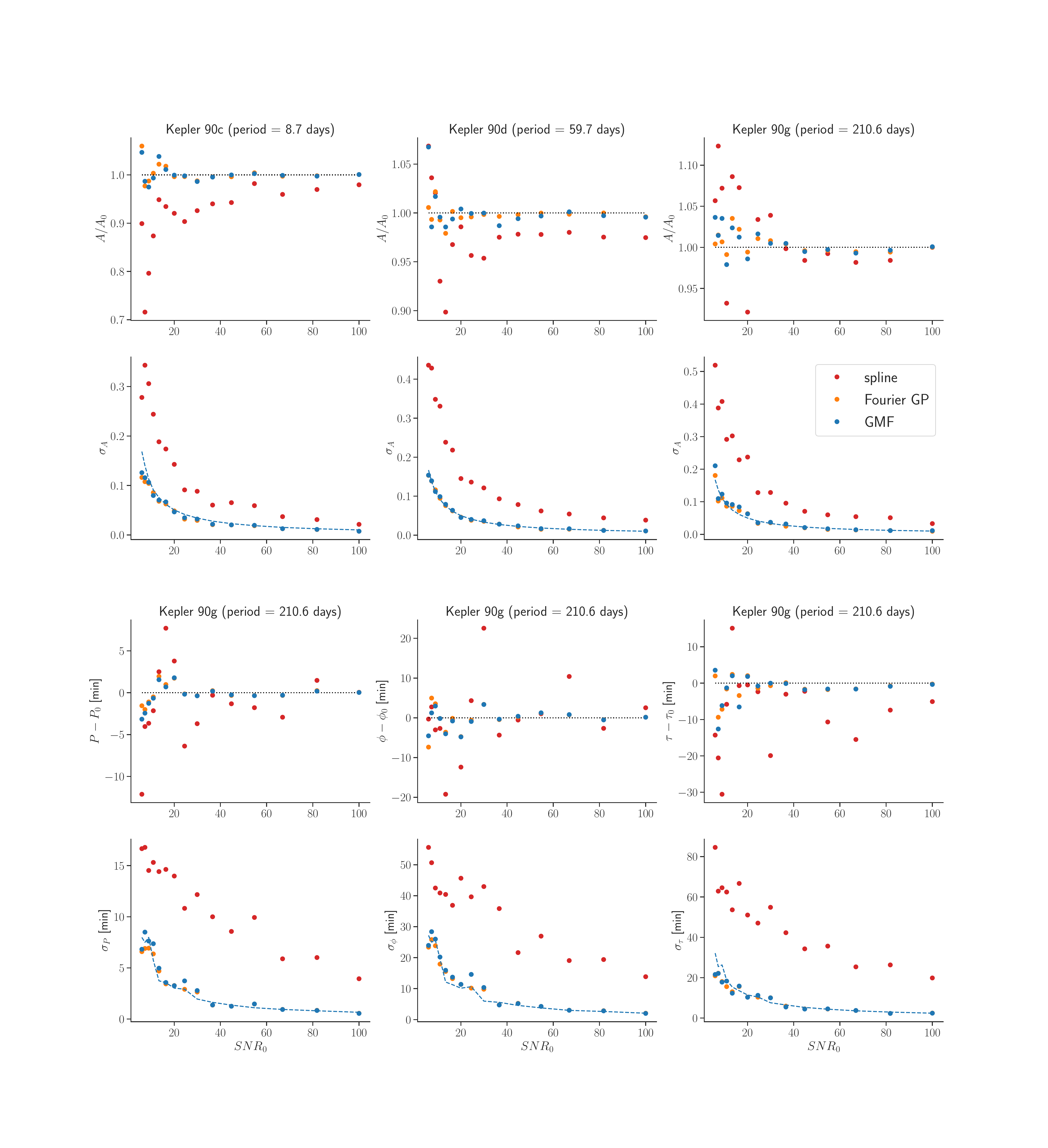} 
\vspace*{-1.5cm}\caption{
{ 
Parameters obtained by the matched filter are compared to the known value of the injected synthetic planets' parameters, as a function of the injected $SNR_0$ in the range 6--100. Upper row shows the expected value of the ratio $A/A_0$ averaged over 30 randomly chosen planet phases to reveal possible bias. Second row shows the variance of  $A$. Three choices for the period of the injected planet are chosen resembling the known planets Kepler 90c, Kepler 90d and Kepler 90g.  The last two rows show the bias and variance of the other three parameters: period, phase and transit duration. In all figures a dotted envelope is a GMF prediction of the error as computed from the Hessian at the $SNR$ peak. It matches well the variances from the simulations.
Matched filter is compared to the Fourier GP and to the spline fitting, showing that is as accurate as FGP and significantly better than the spline fitting in both bias and variance. 
}
}\label{fig:Test}
\end{figure*}

The baseline methods in the 
literature use outlier removal at 12.3 sigma \citep{kepler_process}, which would fall 
between the 7 sigma rejection and no rejection in 
the right hand side of Figure \ref{fig:pvalues}. 

As a result we expect the median of a false 
positive to be around 6-7 for the lower periods below 50 days, increasing to 7-12 for the longer periods. 
Kepler threshold of 7.1 \citep{kepler_process}
may give a low false positive rate
for the shorter periods if spline or wavelets are 
used, but for the longer 
periods it 
may not be sufficiently conservative, as we expect many false positives with $SNR>7$. In contrast, our GMF method achieves median false positive $SNR$ of 5.3 even at long periods: this can lead to a dramatic difference in efficiency of planet detections, specially for Earth like planets in the habitable zone with periods longer than 200 days. At the 
equal false positive rate we expect GMF to 
be sensitive to up to 30\% lower 
transit amplitudes.

\subsection{Planet parameters}
% \begin{figure*}
% \includegraphics[scale = 0.38]{period_test.pdf} 
% \caption{Period obtained by the matched filter $P$ is compared to the known value of the injected synthetic planets $P_0$ as a function of the injected $SNR_0$ analogously to the amplitude test in the Figure \ref{fig:AmplitudeTest}. GMF is completely equivalent to Fourier GP with minimal variance and no bias. Variance is also correctly predicted by the Hessian at the peak.}\label{fig:PeriodTest}
% \end{figure*}
% \begin{figure*}
% \includegraphics[scale = 0.38]{phase_test.pdf} 
% \caption{Temporal phase obtained by the matched filter $\phi$ is compared to the known value of the injected synthetic planets $\phi_0$ as a function of the injected $SNR_0$, analogous to Figure \ref{fig:AmplitudeTest}. Injected phases are chosen randomly for each realization.}\label{fig:PhaseTest}
% \end{figure*}
% \begin{figure*}
% \includegraphics[scale = 0.38]{duration_test.pdf} 
% \caption{Transit duration obtained by the matched filter $\tau$ is compared to the known value of the injected synthetic planets $\tau_0$ as a function of the injected $SNR_0$, analogous to Figure \ref{fig:AmplitudeTest}. Injected transit durations are 4.64 h, 8.81 h and 13.42 h from left to the right, simulating circular, perfectly aligned orbits.}\label{fig:DurationTest}
% \end{figure*}

\begin{figure*}
    \centering
    \hspace*{-0.7cm}\includegraphics[scale = 0.5]{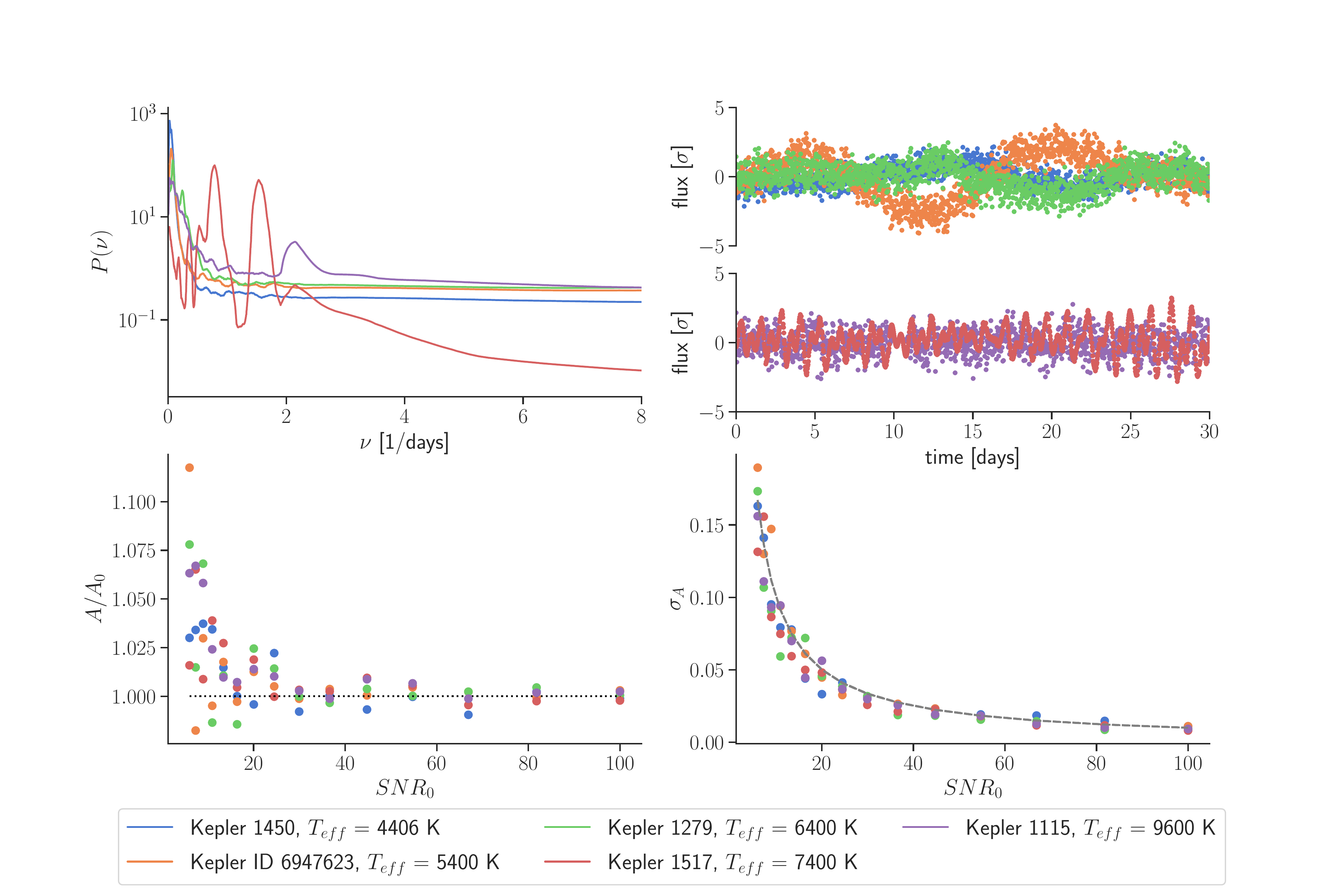}
    \caption{
    {
    We test GMF on a variety of stellar backgrounds. The upper left plot shows the power spectra of the analyzed stars. Note  the large differences in the stellar power spectra, for example Kepler 7515679 (green) has a power spectrum amplitude which ranges over 4 orders of magnitude across the range of frequencies
    shown, while power spectrum of Kepler 5022828 (red) spans only two orders of magnitude. Upper right plot shows a segment of Kepler's measurements for those stars. Time is measured relative to the beginning of Kepler's measurement for that star. Very red spectra (large power at low frequencies compared to the high frequencies), for example Kepler 7515679, are seen as a strongly correlated stellar flux on the timescales shown.
    We take Kepler's measurements, find planets and eliminate them. We then inject a synthetic planet with a period 30 days in the stellar flux and test GMF's ability to extract the injected planet's amplitude $A_0$. Lower left plot shows the expected value of the ratio $A/A_0$ averaged over 30 randomly chosen planet phases as a function of the injected planet's $SNR_0$ in the range 6--100. Lower right plot shows the variance of $A$. A dotted envelope is a GMF prediction of the error as computed from the Hessian at the $SNR$ peak. It matches well the variances from the simulations. GMF's is performing well, regardless of the stellar background. 
    }
    }
    \label{fig: multiple stars}
\end{figure*}

To explore the accuracy of the planet parameter
estimates we inject a planet transit signature with a known $SNR_0$ into the stellar background and test GMF capability to reproduce the injected parameters of the planet: period, phase, transit duration and amplitude. We test our method against the more sophisticated methods Fourier GP \citep{FourierGP} and Celerite GP \citep{celerite}, which are computationally too expensive for a 
planet search, as they need a good initial guess of the planets parameters. The Fourier GP method is argued to be optimal under the assumption of Gaussian stellar variability, so comparing GMF against 
these methods is informative on the (sub)optimality of GMF. 

We show in Figure \ref{fig:Test} that GMF is as good in reproducing the parameters of a planet as the Fourier GP, and is significantly better than the spline fitting, both in terms of bias and variance. The errors equal 
those of Fourier GP, which is near optimal. 
We also show that the matched filter correctly predicts the errors on the planet's parameters. Covariance matrix of the planet's parameters is calculated as an inverse of the Hessian of the $-\log p$, and this analytic method matches the variance obtained from the simulations well. Note the ease of estimating the covariance matrix compared to the GP methods where a marginalization over the stellar parameters is required.

%The Hessian appears noisy and and may underestimate the variance at $SNR<10$, but is accurate above it. 
There is good agreement between GMF and Fourier GP. This shows that GMF can be used for the full analysis, not only for the initial detection analysis, as it gives optimal results on all the parameters of interest. 

{ 
We also test GMF's performance in different stellar backgrounds, to show that it is a general method which can be applied to any star. We take a selection of Kepler's targets and perform the planet search as described in Section \ref{sec: planet search}. We retrieve the known planets and subtract them from the data. We then inject a planet in the data and test GMF's ability to extract its parameters. Results are shown in Figure \ref{fig: multiple stars}. GMF performs well on all tested stars, despite their qualitatively different behaviour.

}

\subsection{Analyzing real planets}

\begin{table*}
\caption{Individual transits of the large planets of Kepler 90 (90g and 90h) for GMF, both transit time, transit duration, and the estimated error. Results from the Fourier GP \citep{FourierGP} are also shown relative to GMF, but the two are not completely comparable, because FGP results also account for the non uniformity of Kepler data point measurements, which is an effect on the order of the error of the parameter. As a consequence, for both transit time and transit duration the agreement between GMF and FGP can be a factor of two larger than the estimated GMF errors.Note that for the transit time a typical difference between the two is of order 3 minutes, while the data are given in 29.4 minute intervals.} 
\label{table:individualPlanets}
\begin{tabular}{|l|c|c|c|c|c|c|}\hline
\bfseries planet & \bfseries SNR & \bfseries transit time [days] &\bfseries transit duration [hours] &\bfseries FGP transit time [days] &\bfseries FGP transit duration [hours]
\begin{filecontents*}{individualPlanets.csv}
planet,SNR,time,transitDuration,FGPtime,FGPtransitDuration
h,108.1,8.96422 $\pm$ 0.0006,13.341 $\pm$ 0.029,0.0018,0.06
g,53.9,15.5869 $\pm$ 0.0011,11.392 $\pm$ 0.055,0.0018,0.07
g,53.9,226.0428 $\pm$ 0.0011,11.349 $\pm$ 0.054,-0.002,0.02
h,109.7,340.60774 $\pm$ 0.0007,13.462 $\pm$ 0.034,0.0005,-0.11
g,53.6,436.7707 $\pm$ 0.0011,11.432 $\pm$ 0.055,0.0025,0.04
g,53.5,857.96 $\pm$ 0.0012,11.621 $\pm$ 0.059,0.0051,-0.0
g,55.4,1068.5496 $\pm$ 0.0025,11.5 $\pm$ 0.12,0.0045,0.01
g,52.8,1280.221 $\pm$ 0.0012,11.676 $\pm$ 0.06,0.0062,-0.04
h,105.9,1335.3784 $\pm$ 0.00058,13.297 $\pm$ 0.028,0.0026,0.06
\end{filecontents*}
\csvreader[head to column names]{individualPlanets.csv}{}
{\\ \hline \planet & \SNR & \time & \transitDuration & \FGPtime & \FGPtransitDuration }
\\\hline
\end{tabular}
\end{table*}

\begin{table*}
\caption{Joint transits of the small Kepler 90 planets from GMF method. We also show $SNR$ of Fourier GP, which agrees well with the $SNR$ of GMF.} \label{table:periodicPlanets}
\begin{tabular}{|l|c|c|c|c|c|c|c|}\hline
\bfseries planet & \bfseries SNR & \bfseries period [days] & \bfseries phase [days] &\bfseries transit duration [hours] & \bfseries FGP SNR
\begin{filecontents*}{periodicPlanets.csv}
planet,SNR,period,phase,transitDuration,SNRfgp
d,27.0,59.73688 $\pm$ 0.00028,27.4461 $\pm$ 0.003,8.003 $\pm$ 0.097,29.6
e,23.1,91.9401 $\pm$ 0.00042,2.7822 $\pm$ 0.0037,8.96 $\pm$ 0.12,23.5
f,17.9,124.915 $\pm$ 0.0013,123.1868 $\pm$ 0.0075,10.08 $\pm$ 0.18,16.6
c,16.2,8.719734 $\pm$ 2.9e-05,8.0097 $\pm$ 0.002,3.831 $\pm$ 0.092,15.8
b,15.4,7.008546 $\pm$ 3.2e-05,6.1096 $\pm$ 0.0033,4.32 $\pm$ 0.13,15.5
\end{filecontents*}
\csvreader[head to column names]{periodicPlanets.csv}{}
{\\ \hline \planet & \SNR & \period & \phase & \transitDuration & \SNRfgp}
\\\hline
\end{tabular}
\end{table*}

{ 
We have applied the GMF to the planet search in several Kepler's targets and have retrieved all officially confirmed planets. }
We show the results for the Kepler 90 system in Table \ref{table:individualPlanets} for the large planets where each transit is identified individually, and in Table \ref{table:periodicPlanets} for the smaller planets which are folded over their period.
In this paper we do not consider the proposed Kepler 90i \citep{neuron_networks}, which is close to the detection threshold, and requires a more careful 
analysis of the Look Elsewhere Effect \citep{LEE1}, which 
we will pursue elsewhere. 
Transit time is defined as the  
time at the center of the transit measured relative to the beginning of the Kepler's measurements in Kepler-90, which is HJD - 2454833 + 131.5124 days. Phase is defined as the transit time of the first observed transit.

We compare these results to those of Fourier GP. 
We again confirm a perfect agreement between 
the GMF and the Fourier GP on estimated $SNR$, see Table \ref{table:periodicPlanets}. 
GMF can be used not only as a planet search 
algorithm, but can also replace expensive GP 
based analysis methods such as Fourier GP or 
Celerite.

\section{Conclusions}
In this paper we propose a method for planet 
detection in transit data, which is based 
on matched filter technique, combined with the 
Gaussianization method for the noise outliers, 
which we call Gaussianized Matched Filter (GMF). 
We show that GMF significantly outperforms
standard baselines in terms of reducing the 
false positive rate: while standard methods 
give median false positive signal to noise 
as high as 8, the corresponding number for GMF 
is 5.3. Since the number of false positives 
explodes exponentially at lower $SNR$ values, 
this could 
enable a significantly lower false positive rate 
of faint planets, specially for Earth 
like planets in the habitable 
zone. Alternatively, at the equal false 
positive rate we expect GMF to detect planets 
with up to 30\% lower transit amplitudes. 

A procedure for GMF planet search can be summarized as: 
\begin{itemize}
    \item Gaussianize the data by remapping the outliers. 
    \item calculate the noise power spectrum and use it in inverse noise weighting in the convolution of the data with the transit profile Fourier transform.
    \item search over the time of transit and transit duration to find the individual transits of the big planets.
    \item search over the period, phase and transit duration for the small planets.
\end{itemize} 

The method eliminates outlier false positves and stellar variability 
positives that in the standard Kepler pipeline 
need to be eliminated in the post-processing 
phase using RoboVetter \citep{robovetter}. 
Further false positives that need to be considered are binary stars \citep{foremanplanetsearch} and off-target false positives, and we plan to address these elsewhere. 

A remarkable feature of the GMF method is that 
it can be used not only for the initial planet 
detection, but also for the final planet parameter analysis. By comparison against the state of the art GP methods on both simulations and on 
real data we observe that GMF achieves 
near optimal results on amplitude, period, phase, and transit time. Moreover, a simple analytic 
Laplace approximation of the Hessian gives 
reliable error estimates on these parameters.
Thus GMF may be 
not only the fastest method to detect 
planets, with the lowest false positive rate, 
but also the most accurate method 
to extract their parameters.

GMF provides parameter estimates and their errors as a compressed summary statistic, 
 from which one can form a likelihood that one can 
 use for the more involved inverse problems, 
where optimization or  
Monte Carlo Markov Chain (MCMC) analysis is 
needed to find the solution. 
{
One such example is a transit timing 
variation (TTV) analysis \citep{yanTTV}, where transit
times and transit durations and their errors 
from the GMF analysis of 
individual transits in Table \ref{table:individualPlanets} have been used 
to form a data likelihood. This enabled a subsequent inverse problem analysis that 
identified the models that can explain TTVs. 
In this example this led to a determination of 
all of the orbital parameters and the masses of 
Kepler 90g and h, and the discovery of Kepler 90g
as a superpuff.
}

\section*{Acknowledgements}
We acknowledge Ad futura Slovenia for supporting J.R. MSc study at ETH Zürich. This material is based upon work supported by the National Science Foundation under Grant Numbers 1814370 and NSF 1839217, and by NASA under Grant Number 80NSSC18K1274. {This paper includes data collected by the Kepler mission. Funding for the Kepler mission is provided by the NASA Science Mission directorate.}

\section*{Data Availability}
The data underlying this article are available in NASA Exoplanet Archive, at  \url{https://exoplanetarchive.ipac.caltech.edu/bulk_data_download/}.

\bibliographystyle{mnras}
\bibliography{cosmo,citations}

%\bsp	% typesetting comment
%\label{lastpage}
\end{document}